\documentclass[prb,twocolumn,superscriptaddress]{revtex4}
\usepackage[dvips]{graphicx}
\usepackage{latexsym}
\usepackage{amsmath}
\begin{document}
%\preprint{cond-mat/0305xxx}
\title{Finite temperature properties of quantum Lifshitz transitions between valence bond solid phases:
An example of `local' quantum criticality}
\author{Pouyan Ghaemi} \affiliation{Department of Physics, Massachusetts
  Institute of Technology, Cambridge MA 02139}
\author{Ashvin Vishwanath} \affiliation{Department of Physics,
  University of California, Berkeley, CA 94720.}
\author{T. Senthil} \affiliation{Department of Physics, Massachusetts
  Institute of Technology, Cambridge MA 02139}
\date{\today}
\begin{abstract}

We study the finite temperature properties of quantum magnets close to
a continuous quantum phase transition between two distinct valence
bond solid phases in two spatial dimension. Previous work has shown
that such a second order quantum `Lifshitz' transition is described
by a free field theory and is hence tractable, but is nevertheless
non-trivial. At $T>0$, we show that while correlation functions of
certain operators exhibit $\omega/T$ scaling, they do not show
analogous scaling in space. In particular, in the scaling limit, all
such correlators are purely {\em local} in space, although the same
correlators at
$T=0$ decay as a power law.  This
provides a valuable microscopic example of a certain kind of `local'
quantum criticality. The local form of the correlations arise from the
large density of soft modes present near the transition that are
excited by temperature. We calculate exactly the autocorrelation
function for such operators in the scaling limit. Going beyond the
scaling limit by including irrelevant operators leads to finite
spatial correlations which are also obtained.

\end{abstract}
\newcommand{\fig}[2]{\includegraphics[width=#1]{#2}}
\newcommand{\be}{\begin{equation}}
\newcommand{\ee}{\end{equation}}
\maketitle
\section{Introduction}
Recent theoretical work \cite{sc} has discussed strange and unusual
phenomena in the vicinity of certain quantum phase transitions in
insulating magnets and related systems. These phenomena -dubbed
`deconfined quantum criticality' - do not fit in easily into the
Landau-Ginzburg-Wilson paradigm for phase transitions. Such quantum
critical points seem to be most aptly described in terms of
fractionalized degrees of freedom that interact through emergent
gauge interactions. These fractionalized modes are in general absent
in the phases on either side of the transition but rear their head
right at the critical point.

Perhaps the most analytically tractable example of such a deconfined
quantum critical point (or indeed of {\em any} interesting quantum
critical point in dimensions bigger than one) is provided by a phase
transition between two different valence bond solid phases of
spin-$1/2$ quantum Heisenberg magnets that was discussed in
Ref.~\onlinecite{huse,VBS}.  A specific example of such a quantum
`Lifshitz' transition is a phase transition between a featureless
valence bond solid and a different translation broken one on a
bilayer honeycomb lattice. Closely related are the better studied
transitions that occur at the Rokhsar-Kivelson(RK)
points\cite{henley} of quantum dimer models on bipartite lattices.
For instance on the square or honeycomb lattices, the RK point
separates two ordered conventional phases. Unfortunately the RK
point corresponds to a special very fine tuned multicritical point.
(This difficulty is however not present for the bilayer honeycomb
model). The tractability of these quantum phase transitions arises
from the existence of a free field description. Despite this the
theory has non-trivial structure as shown in Refs.
\onlinecite{VBS,huse,Ardonne}. A close mathematical analogy may be drawn
with non-trivial critical points in $1+1$ dimensions which too have
free field descriptions, and are hence tractable. The free field
description of the two dimensional critical points of interest in
this paper may be given either in terms of a (non-compact) $U(1)$
gauge theory or equivalently in terms of its dual sine-Gordon theory.
The latter may be more familiar to readers
conversant with `height' descriptions of quantum dimer models\cite{ReSaSuN,Frbook} and will be
used through out this paper. In this height description the critical
theory takes the form:
\begin{equation}
S_0 = \frac{1}{2} \int d \tau d^2 x \{\left(\partial_{\tau} \chi
\right)^2 + K \left(\nabla^2 \chi \right)^2\}
\end{equation}
Here $\chi$ is the height field. This theory actually describes a
fixed line that is parameterized by $K$\cite{Grinstein}.
Further details may be found in Ref.~\onlinecite{VBS,huse,Ardonne}. We
first note that the form of the critical action immediately implies
that the dynamic critical exponent $z = 2$. Various physical
observable have non-trivial scaling structure at this critical point
at zero temperature. The purpose of the present paper is to focus on
finite temperature correlators of this critical theory - in
particular for dynamical correlators. The free field form makes
these calculations feasible - a unique property of this quantum
transition among other non-trivial two dimensional ones.

Of particular interest to us will be operators such as
$\exp{(2i\pi\chi)}$. Such operators correspond in the gauge theory
interpretation to monopole or instanton events which change the
total gauge flux of a state by $2\pi$. In the context of quantum
dimer models at their RK points, these operators have another
interpretation\cite{ReSaSuN}. They are simply the order parameters for one of the
phases (the columnar/plaquette) on one side of the transition. As
expected, their correlators decay as power laws in both space and
time at zero temperature. In this paper we calculate the correlators
of such operators at finite temperature. Remarkably in the scaling
limit we show that these have the striking property of being {\em
short ranged} in space. The autocorrelation (at two different times
at the same spatial point) is however non-zero and non-trivial. This
remarkable property implies that at a non-zero temperature $T$ the
corresponding frequency ($\omega$)  and momentum ($k$) dependent
susceptibility shows $\omega / T$ scaling but has no momentum
dependence! We show these features by exact calculation of the
universal scaling function for the susceptibility. The unusual
structure found in the scaling limit suggests that any spatial
structure in the correlator is due to formally irrelevant terms that
correct the leading scaling behavior. We determine the form of these
corrections to scaling that restore spatial correlations.

The theoretical phenomena found in this paper share some
similarities with various ideas that have been proposed in theories
of many interesting correlated materials. Indeed $\omega/T$ scaling
is seen in experiments on a number of different systems - most
prominently in the normal state of optimally doped cuprates and in
the non-fermi liquid metals near heavy electron critical points.  In
general, evidence for scaling phenomena in spatial correlations is
much weaker (for instance in the cuprates). An interesting viewpoint
on such scaling in various strange metals is to attribute it to
universal singularities of some proximate critical point. A number
of workers have advocated some kind of spatially localized
fluctuations at the relevant quantum critical points to account for
the weaker signatures of scaling in spatial correlations\cite{varma,si}.  However
the theoretical framework for description of such exotic critical
points (should they even exist) is unclear at present.  The model
described in the present paper may perhaps shed some light on such
theoretical issues (see Section \ref{conc}).

\section{Brief Review of Zero Temperature Criticality and Lattice Effects}

The critical theory associated with the zero temperature Lifshitz
transition between VBS phases has been discussed in detail elsewhere
\cite{huse, VBS}. Here we simply note that under the appropriate
conditions (eg. a bilayer honeycomb spin system with
antiferromagnetic interlayer exchange) a continuous transition is
possible, from a VBS state with zero `tilt' to one where the tilt
begins to grow. While the details of the physics on the tilted side
is complicated \cite{huse}, here we will restrict ourselves  to
reviewing briefly properties of the zero temperature transition
itself. It is convenient to use the height representation of the VBS
states, in terms of which the effective action (eg. for the bilayer
honeycomb model) takes the form:
\begin{eqnarray}
\label{sgcont}
S & = & S_0 + S_1 + S_{inst} \\
S_0 & = & \frac12 \int \! d^2x d\tau  \big\{ (\partial_\tau \chi)^2 + \rho (
{\bf \nabla}  \chi)^2
+K (\nabla^2 \chi)^2\big\}\\
S_1 & = & \int \! d^2x d\tau \, \frac u 4 |{\bf \Delta}\chi|^4 + \dots \\
S_{inst} & = & -\int \! d^2x d\tau \, \lambda \cos(2\pi \chi)
\end{eqnarray}
where the height field $\chi$ plays the role of a dual gauge
potential and is related to the electric field strengths of the
gauge theoretic description via $E_i=\epsilon_{ij}\partial_j\chi$.
In this interpretation the $\lambda$ term corresponds to monopole
tunneling events, and the transition of interests occurs as we tune
the $\rho$ term through zero. At the critical point, the $u$ term is
marginally irrelevant, and the monopole tunneling events are also
irrelevant for some range of $K$ ($0<K<(\frac\pi 8)^2$). The
critical action in the scaling limit is simply given by:
\begin{equation}
S_c=  \frac12 \int \! d^2x d\tau  \big\{ (\partial_\tau \chi)^2 +K
(\nabla^2 \chi)^2\big\}
\label{scalingaction}
\end{equation}
In addition there are logarithmic corrections to correlation
functions arising from the marginally irrelevant term. Although this critical action is
Gaussian, there exist operators in this theory that have nontrivial
scaling dimensions. In particular, we will focus on the monopole
insertion operator $V^\dag(r,\tau) = e^{2\pi i \chi(r,\tau)}$. It
may be easily seen that in the scaling limit, correlations
($C_0(r,\tau)=\langle {\mathcal T}_\tau V^\dag(r,\tau)V(0,0)
\rangle$) of this operator have a non-trivial power law behaviour:
\begin{eqnarray}
C_0(r,0) &\sim&  r^{-\frac\pi{\sqrt{K}}}\\
C_0(0,\tau) &\sim& \tau^{-\frac\pi{2\sqrt{K}}}
\end{eqnarray}
In fact the correlation function can be written in the scaling form
$C_0(r,\tau) \sim r^{-\frac{\pi}{\sqrt{K}}}  f(r^2/\tau)$, which is
a result of the dynamic critical exponent $z=2$ in this theory. For
completeness we note that the scaling function is found to be:
\begin{eqnarray}
f(z) &=& e^{-\frac\pi{2 \sqrt{K}}\Gamma(0,z)}\\
\Gamma(0,z) &=& \int^\infty_z \! dt\frac{e^{-t}}{t}
\end{eqnarray}
We now consider the form of these correlators at finite but small temperatures, where the physics is expected to be controlled by the zero temperature quantum critical point.

\section{Calculation of  Correlations at Finite Temperatures}

Before proceeding to an explicit calculation of the correlators at
finite temperature, we first note what we might naively expect such
a calculation to produce. Consider correlations of monopole
insertion operators at a finite temperature $T>0$ ($C_T(r,\tau)$).
In the quantum critical regime the inverse temperature just provides
a cutoff in imaginary time and is also expected to induce a spatial
length scale below which correlations look quantum critical.
Moreover, one may expect that for an operator with a nontrivial
scaling dimension, finite temperature scaling forms for the
correlation function, in frequency space, may be written as:
\be
\tilde{C}_T(r,\omega) =\frac1T r^{-\frac\pi{\sqrt{K}}}
F(r^2T,\omega/T ) \ee
In particular, the autocorrelation
function may be expected to have the scaling form: \be
\tilde{C}_T(0,\omega) = \frac1T \omega^{\frac\pi{2\sqrt{K}}}
g_1(\omega/T ) \label{auto} \ee while the equal time correlation
function should obey: \be C_T(r,\tau=0)
=r^{-\frac\pi{\sqrt{K}}} g_2(r\sqrt{T}) \label{equaltime} \ee
Further one might expect logarithmic violations of these scaling forms if the marginally irrelevant quartic term is allowed
in the microscopic model.
Below
we will show that while the autocorrelation function does exhibit
the $\omega/T$ scaling form shown in Eqn. (\ref{auto}), the equal
time correlation function at spatially separated points  does {\em
not} exhibit the expected scaling form of Eqn. (\ref{equaltime})!
Moreover, in the scaling limit the correlation function in Eqn.
(\ref{equaltime}) vanishes at any two spatially separated points due
to an infrared divergence. In other words the correlations in the
scaling limit are purely local. This is demonstrated in the two
sections below. In order to obtain a non-vanishing part for the
spatial correlators we have to go beyond the scaling limit and
include the effect of operators that are irrelevant at the zero
temperature critical point. In situations in which the quartic $u$ term is allowed its marginal irrelevance leads to
logarithmic violations of scaling. We will show that non-vanishing spatial correlations result entirely from these
logarithmic corrections to the naive scaling limit.
These effects, as well as the effects
arising from gapped spinons will be discussed in the third section
below.
\subsection{Calculations in the Scaling Limit}
\subsubsection{Autocorrelation Function}

In this subsection we calculate the autocorrelation function at finite
temperatures of the monopole insertion operator. The spectral function
corresponding to this autocorrelation function in the scaling limit
can be calculated exactly and is displayed in Eqn. (\ref{A}). In the
following we describe the details of that calculation.

To calculate finite temperature properties near the
quantum critical point in the scaling limit, we use the fixed point
Euclidean action:
\begin{equation}
\label{action} S_{c}=\int_{0}^{\frac{1}{T}} d\tau \int d^2x
\frac{1}{2} \{(\partial_{\tau}\chi)^{2}+ K(\nabla^{2}\chi)^{2}\}
\end{equation}
where $T$ is the temperature. The free field nature of this action
allows us to readily compute correlation functions.The marginally
irrelevant quartic term has been dropped in this subsection - that will
only lead to logarithmic corrections to the results derived below,
and hence will be ignored in what follows. However it will play an important role in the structure of the
spatial correlations and will be reinstated in Section \ref{qrtsec}.

We begin with a calculation of the autocorrelation function in the scaling limit at non-zero temperature. This is defined to be
\begin{equation}\label{auto}
C_{0}(0,\tau) = \langle
 e^{i2\pi\chi_{0}(\tau)}e^{-i2\pi\chi_{0}(0)}\rangle
\end{equation}
For the Gaussian action this is readily evaluated
at finite temperature and takes the form:
\begin{equation}
C_{T}(0,\tau) = \exp(-T \sum_{n=-\infty}^{\infty}
(1-e^{i\omega_{n} \tau })\int \frac{d^{2} q}{\omega_{n}^{2}+K
q^4})
\end{equation}
After integrating over $\vec{q}$ we get:
\begin{equation}
C_{T}(0,\tau) = \exp(-\frac{\pi}{2 \sqrt{K}} \sum_{n=1}^{\infty}
\frac{1-\cos(2 \pi n \tau T)}{n})
\end{equation}
Performing the sum over $n$ we have \cite{footnotesum}:
\begin{equation}\label{auto-corr}
C_{T}(0,\tau)  =
\frac{c_0T^{\eta}}{(1-\cos {(2\pi \tau T)} )^{\frac{\eta}{2}}}
\end{equation}
where
\be
\eta = \frac{\pi}{2 \sqrt{K}}
\ee
and $c_0$ is a constant. Clearly, in the zero temperature limit this
reduces to equation (\ref{auto}), while at any finite temperature it
has a scaling form (i.e. $T^{-\eta}C_T(0,\tau)$ is only a function of
the product $T\tau$).  Thus, the autocorrelation function exhibits the
usual scaling form.

We will proceed below to explicitly calculate the finite temperature
spectral function associated with this correlator, since it is one
of the the few situations in 2+1 dimensional criticality where this
may be done. We will follow closely an almost identical calculation
of the finite temperature spectral function of a one dimensional
Luttinger liquid in Ref. \cite{sss}. We begin by taking the Fourier
transform of this imaginary time auto correlation function:
\begin{equation}\label{fourier}
\tilde{C}_{T}(0,i\omega_n)= \frac{c_0T^{\eta-1}\
\Gamma(\frac{1-\eta}{2})\
\Gamma(\frac{\eta}{2}+\frac{|\omega_{n}|}{2 \pi T} ) } { \Gamma(\frac{\eta}{2})\
\Gamma(1-(\frac{\eta}{2}-\frac{|\omega_{n}|}{2 \pi T}))}
\end{equation}
Analytic continuation to real  frequencies is now easily performed. In order to obtain the retarded Green's function we need to approach the real frequency axis from the upper half plane, i.e. perform the replacement $|\omega_n| \rightarrow -i\omega$.  This yields:
\begin{equation}
\tilde{C}^{\rm Ret}_T(0,\omega)=\frac{c_0T^{\eta-1}\ \Gamma(\frac{1-\eta}{2})\
|\Gamma(\frac{\eta}{2}-\frac{i \omega}{2 \pi T})|^{2}}{2^{^{\eta}
\slash_{2}} \ \pi^{^{3} \slash_{2}} \ \Gamma(\frac{\eta}{2})}\
\sin(\frac{\pi \eta}{2} +\frac{i\omega}{2T})
\end{equation}
where we have used the well known identity $1/\Gamma(\frac12 - z)=\frac1\pi\Gamma(\frac12 + z)\cos \pi z$. The spectral function associated with this autocorrelator is then given by:
\begin{eqnarray}
A_T(\omega)&=&2\ {\rm Im}(\tilde{C}^{\rm Ret}_{T}(0,\omega)) \\
    &=&c_0T^{\eta-1} \frac{ \sinh(\frac{\omega}{2
T}) \ |\Gamma(\frac{\eta}{2}-\frac{i\omega}{2\pi
T})|^{2}}{2^{\frac{\eta-2}{2}}\sqrt{\pi}\  \Gamma(\frac{\eta}{2})\
\Gamma(\frac{\eta +1}{2})} \label{A}
\end{eqnarray}
We note that the above spectral function \cite{footnote1} is an odd
function of $\omega$ and  writing it in the form:
\be
A_T(\omega) =c_0T^{\eta-1}F(\omega/T)
\label{scalingdefn}
\ee
we see that it satisfies $\omega/T$
scaling.

\begin{figure}
\label{sfn}
\includegraphics[width=8.7cm]{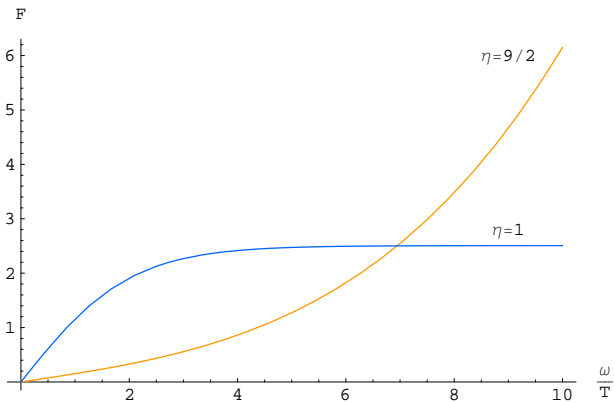}
\vspace{0.1in}
\caption{The scaling part of the spectral function, defined in Eqn
(\ref{scalingdefn})
associated with the
monopole operator autocorrelation function plotted for two different
values of $K$ (and hence $\eta$, the scaling dimension of the monopole
operator). The value $\eta=1$ corresponds to the scaling dimension of
the monopole operator at the square lattice RK point. The value
$\eta=9/2$ corresponds to a point on the stable fixed line that controls the transition
between VBS phases of a bilayer honeycomb quantum magnet.}
\end{figure}

The asymptotic behaviours   of the scaling function are as follows.
For small frequencies, the function must be analytic at finite
temperatures and hence  $F(x) \sim x$ for $x \ll 1$ while at high
frequencies we find $F(x) \sim x^{\eta -1}$ which yield the expected
zero temperature form of the spectral function $A(\omega) \sim
\omega^{\eta -1}$.   Finally we note that for $\eta=1$ which is the
exponent relevant for the square lattice RK point, the scaling
function takes on a particularly simple form: \be
F(\frac{\omega}{T})=\sqrt{\frac8\pi}\tanh(\frac{\omega}{2T})
\label{RKscalingfn} \ee This, along with the scaling function for
$\eta=9/2$ (which corresponds to a value of $K=\pi^2/81$ that is on
the stable fixed line for the bilayer honeycomb model) is plotted in
Fig. \ref{sfn}.

Thus the free Gaussian action has given us a unique opportunity to extract exact information on the
non-trivial real time dynamics at non-zero temperatures above a $2+1$ dimensional quantum critical
point.

\subsubsection{Equal Time Spatial Correlations}
\label{spcorrsc}

 We now proceed to  calculate
the finite temperature, equal-time correlator of the monopole
insertion operator:
\begin{equation}
 C_{T}(r,0) = \langle
e^{i2\pi\chi_{r}(0)}e^{-i2\pi\chi_{0}(0)}\rangle_{T} \label{Tcorr}
\end{equation}
Using the form (\ref{action}) for the action we get the following
expression for the correlator:
 \begin{equation}\label{correlate}
 C_{T}(r,0) = \exp{\{\  -T\int d^{2}q
 \sum_{n=-\infty}^{\infty}\frac{(1-e^{i\vec{q}.\vec{r}})}{\omega_{n}^{2}+Kq^{4}}
\}}
 \end{equation}
 The sum over the Matsubara frequencies ($\omega_{n}=2 \pi n T$) is easily performed:
 \begin{equation}\label{sum}
 T\sum_{\omega_{n}}\frac{1}{\omega_{n}^{2}+Kq^{4}}=\int
 \frac{d\omega}{2\pi}\frac{1}{\omega^{2}+Kq^{4}} +
 \frac{1}{\sqrt{K}q^{2}}\frac{1}{e^{\frac{ \sqrt{K}q^{2}}{T}}-1}
 \end{equation}
 Putting this back into equation (\ref{correlate}), we see that the
 first term on the right hand side of (\ref{sum}) generates the zero temperature
 correlation function. Thus:
 \begin{equation}
 C_{T}(r,0)=C_{0}(r,0)\Phi_{T}(r)
 \end{equation}
 Where $C_{0}(r,0)\sim \frac{1}{r^{\frac{\pi}{\sqrt{K}}}}$ is the zero temperature correlator and
 \begin{equation}
\Phi_{T}(r) = \exp(-\int \frac{d\theta dq}{\sqrt{K} q}
 \frac{1-e^{iqr\cos{\theta}}}{e^{\frac{\sqrt{K}q^{2}}{T}}-1})
 \end{equation}
Note however that the integration over $q$ diverges logarithmically at its lower limit. Introducing an infrared cutoff ($1/L$) which is the inverse of the linear dimension of the  system, we have:
\begin{eqnarray}
C_{T}(r,0) &\sim&
\frac{1}{r^{\frac{\pi}{\sqrt{K}}}} e^{-\frac{\pi }{2 K}T r^{2}\log
L}\\
\label{locality}
\nonumber
&\rightarrow& 0 \, \, (L\rightarrow\infty,\, r \neq 0)
\end{eqnarray}

Thus, at any finite temperature, correlations at spatially separated points vanish in the thermodynamic limit ($L\rightarrow \infty$).  Note, that if the zero temperature limit is taken first, then we recover the zero  temperature correlator. Thus, finite temperature correlations (both equal and non-equal time) of this operator in the scaling limit vanish at spatially separated points. In other words the correlations are purely  {\em local}. In fact it is easy to pinpoint the origin of this divergence. The Matsubara sum in (\ref{correlate}) is dominated by the zero frequency contribution which implies that we must examine the integral:
$$
T\int d^2q \frac{(1-e^{i\vec{q}\cdot\vec{r}})}{Kq^4}
$$
which is clearly logarithmically divergent at small $q$ and hence leads to the result in equation (\ref{locality}).  This divergence arises from the large number of soft modes present at low energies in this system, due to the quadratic dispersion. The physical origin of this large number of low lying states is a consequence of the fixed point action (\ref{action}) being symmetric under arbitrary translations $\chi \rightarrow \chi+ {\rm const.}$ {\em and} arbitrary tilts $\chi \rightarrow \chi + \vec{Q}\cdot\vec{r}$  (for any  $\vec{Q}$) of the height field.
Since the scaling limit contribution to the correlation function vanishes at spatially separated points, we need to go beyond the scaling limit in order
to obtain a finite contribution.

\subsection{Beyond the Scaling Limit: Effect of Irrelevant Operators and Gapped Spinons}
\label{beyondscaling}

\subsubsection{ Effect of the Quartic Operator}
\label{qrtsec}

So far we have been concerned with the finite temperature properties
of the quantum critical system in the scaling limit, where all
operators that are irrelevant at the quantum critical point were
omitted. In particular the quartic term, which is marginally
irrelevant at the quantum critical point, and leads only to
logarithmic corrections to correlators, was dropped.
%However, at
%finite temperature the system eventually flows away from the zero
%temperature quantum critical fixed point  to a different (infinite
%temperature) fixed point.
We now discuss the fate of the spatial correlations at non-zero
temperature once the quartic term is included. Subsequently we
discuss the effect of thermally excited spinons (vortices in the
height field). The crucial effect is that a non-zero value is generated for the strength of the quadratic spatial gradiant term
(denoted $\rho$ in Eqn. \ref{sgcont} above) - this is familiar in the analysis of quantum criticality for models at or
above their upper critical dimension\cite{subir}. Below we show how this restores spatial correlations for the monopole operators discussed above.

We begin by calculating the value of $\rho$ at low non-zero temperatures above the quantum
critical point. This may be done by considering the renormalization group flows at $T \neq 0$ away from the zero temperature critical fixed point.
Equivalent results are obtained in a simple approximation that is exact for a suitable large-$N$ generalization of the model.
We will present this calculation below.

 Retaining the quartic term in our
continuum effective action we have:
\begin{equation}\label{act2}\begin{split}
S=\frac{1}{2}\int_{0}^{\frac{1}{T}}d\tau \int d^{2}x\
\{&(\partial_{\tau}\chi)^{2}+\rho^{QC} (\nabla \chi)^{2}\\
&+K(\nabla^{2}\chi)^{2}+u(\nabla\chi)^{4}\}
\end{split}
\end{equation}
Note that we have included a bare stiffness term $\rho^{QC}$ which
is determined by requiring that at zero temperature the system is at
the quantum critical point.

In order to handle the quartic term, we will resort to the large N
approximation, where we will assume that $\vec{\chi}$ is an $N$
component field. Of course, we are strictly interested in the $N=1$
limit, but it will be convenient to consider a large number of components.
%assuming that $\chi$ filed is a $N$ component field,
 In order to obtain a sensible action in this limit, we rescale $u$ to $u/N$ so that the quartic term
in the action takes the form $\frac{u}{N}|\nabla \vec{\chi}|^4$. We can
rewrite the partition function as:
\begin{equation}
Z=\int [D\chi] [D\lambda] \ e^{-S[\vec{\chi},\lambda]}
\end{equation}
where
\begin{equation}
\begin{split}
S= \int d\tau d^2x
[\frac{|\partial_\tau\vec{\chi}|^2}{2}+&K\frac{|\nabla^2\vec{\chi}|^2}{2}\\
+&\rho^{QC}\frac{|\nabla\vec{\chi}|^2}{2}+i\lambda\frac{|\nabla
\vec{\chi}|^2}{2}+\frac{N\lambda^2}{16 u}]
\end{split}
\end{equation}
Now replacing $i\lambda$ with $\rho_{eff}-\rho^{QC}$,  (this choice
will simplify notation) we have a path integral over the fields
$\vec{\chi}$ and $\rho_{eff}$. Now preforming the path integration
over $\vec{\chi}$ we obtain:
\begin{equation}
Z\sim\int [D\rho_{eff}]\ e^{-S[\rho_{eff}]}
\end{equation}
where:
\begin{equation}\begin{split}
S[\rho_{eff}]=&\frac{NL^2}{2}\int \frac{d^2q}{(2\pi)^2}
\sum_{\omega_n} \log(\omega_n^2+Kq^4+\rho_{eff}\
q^2)\\&-\frac{N(\rho_{eff}-\rho^{QC})^2}{16 u}\beta L^2
\end{split}
\end{equation}
here $\omega_n=2\pi nT$ are Matsubara frequencies. We see that
$S[\rho_{eff}]$ is of order $N$ and so in the large $N$ limit, we
can preform the integration over $\rho_{eff}$ using \textit{saddle
point} method. This gives the following self consistent equation for
$\rho_{eff}$:
\begin{equation}\label{self}\begin{split}
\rho_{eff}-\rho^{QC}=&4uT\int
\frac{d^2q}{(2\pi)^2}\sum_{\omega_n}\frac{q^2}{\omega_n^2
+Kq^4+\rho_{eff}q^2}\\
=4u&\int
\frac{d^2q}{(2\pi)^2}\frac{q^2}{\sqrt{Kq^4+\rho_{eff}q^2}}(\frac{1}{2}+\frac{1}{e^{\frac{\sqrt{Kq^4+\rho_{eff}q^2}}{T}}-1})
\end{split}
\end{equation}

where we have used the identity:\be\nonumber
T\sum_{\omega_n}\frac1{\omega_n^2 + a^2}=\frac1a \left[ \frac12 +
\frac1{e^{\beta a}-1}\right]\ee

The bare parameter $\rho^{QC}$ is calculated by requiring that
at zero temperature the system is at the quantum critical point, i.e. $\rho_{eff}=0$.  Setting
$\rho_{eff}\rightarrow 0$ as $T\rightarrow 0$ in equation
(\ref{self}), we get:
\begin{equation}
-\rho^{QC}=4u\int \frac{d^2q}{(2\pi)^2}\frac{q^2}{2\sqrt{Kq^4}}
\end{equation}
Putting all these together we get the following self consistent
equation in which the only unknown parameter is $\rho_{eff}$:
\begin{equation}\begin{split}
\rho_{eff}=4u\int\frac{d^2q}{(2\pi)^2} \{&\
 \frac{q^2}{\sqrt{Kq^4+\rho_{eff}q^2}}\\&(\frac{1}{2}+\frac{1}{e^{\frac{\sqrt{Kq^4+\rho_{eff}q^2}}{T}}-1})
-\frac{1}{2\sqrt{K}}\}
\end{split}\end{equation}
This integral is preformed assuming
a high momentum cutoff  $\Lambda$.
Assuming we are in the regime where $\rho_{eff}\ll \sqrt{K}T$ we
get:
\begin{equation}\label{eff}
\rho_{eff}=\frac{uT}{\pi K}\
\frac{\log{(\frac{\sqrt{K}T}{\rho_{eff}})}}{1+\frac{u}{4\pi
K^{\frac{3}{2}}}[\log{(\frac{4K\Lambda^2}{\rho_{eff}})}-1]}
\end{equation}
In the limit of low $T \rightarrow 0 $, we expect $\rho_{eff} \rightarrow 0$ so that the denominator may be approximated by keeping only the
logarithm. This gives the following self-consistency equation.
 \begin{equation}
\rho_{eff} \approx 4T\sqrt{K}
\frac{\log(\frac{\sqrt{K}T}{\rho_{eff}})}{\log(\frac{4K\Lambda^2}{\rho_{eff}})}
\end{equation}
This may be solved to give (in the low $T$ limit)
\begin{equation}\label{rho}
\rho_{eff} \approx 4T\sqrt{K}\
\frac{\log{(\log{(\frac{1}{T})})}}{\log{(\frac{1}{T})}}
\end{equation}

Equation (\ref{rho}) shows that for small $T$,
$\rho_{eff}\ll\sqrt{K}T$. This justifies our assumption in deriving
equation(\ref{eff}). Also note that $\rho_{eff}$ goes to zero as $T$
goes to zero. Naive scaling based on the dynamical critical exponent $z = 2$ would have suggested $\rho_{eff} \sim T$.
This is violated by logarithmic corrections which is exactly what is expected at the upper critical dimension due to the
marginal irrelevance of the $u$ term.

Now with this $\rho_{eff}$ we have the following effective action for finite temperature:
\begin{equation}
S=\frac{T}{2}\int d^2x\{K(\nabla^2\chi)^2+\rho_{eff}(\nabla\chi)^2\}
\end{equation}

Therefore in this limit the equal time correlation function takes
the form: \be C_T(r) \approx \langle e^{i 2\pi T ({\chi}(r)
-{\chi}(0))}\rangle = e^{-\Phi_T(r)}\label{corr} \ee where \be
\Phi_T(r)=T\int d^2q \frac{1-e^{iq\cdot r}}{Kq^4 + \rho_{eff}q^2}
\ee
 Performing the angular integral above and rewriting in terms of
the scaled variables $k=|q||r|$ and introducing the characteristic
length scale $\xi_T$:
\be
\xi_T^2=K/\rho_{eff}
\label{xi_T}
\ee

we have

\be
\begin{split}
\Phi_T(r)&=\frac{2\pi Tr^2}{K}\int _0^\infty \frac{k dk
(1-J_0(k))}{k^4
+ \frac{r^2}{\xi_T^2}k^2}\\
&=\frac{2\pi T\xi_T^2}{K}[K_0(r/\xi_T)+\log(r/\xi_T) + C - \log 2]
\end{split}
\ee where $J_0$ and $K_0$ are Bessel functions, and $C=0.5772...$ is
the Euler constant. The asymptotic  properties of this function are
as follows. First let us consider $r \ll \xi_T$: \be \Phi_T(r \ll
\xi_T) = \frac{\pi T r^2}{2K}[\log(2\xi_T/r)+(1-C)] \ee
This is
essentially the correlation function we obtained in the scaling
limit (\ref{locality}) with $\xi_T$ playing the role of the system
size and cutting off the divergent integral. This is as it should
be; the length scale $\xi_T$ represents the crossover scale from the
scaling behaviour at shorter scales to the behaviour that is
characteristic of the eventual finite temperature phase at larger
scales. The behaviour at these larger scales can be obtained by
studying the $r \gg \xi_T$ behaviour of the function above:
\be
\Phi_T(r \gg \xi_T) = \frac{2\pi T}{\rho_{eff}}(\log(r/\xi_T)+C)
\ee

Thus, the form of the correlation function at finite temperatures at
the longest scales is simply a power law $C_T(r) \sim 1/r^\sigma$ with
an exponent $\sigma=2\pi T/\rho_{eff}$ that diverges logarithmically
as $T \rightarrow 0$.  The divergence of the exponent at low-$T$
implies a rapidly decaying power law form.

Several comments are in order about this result.  First the
correlation is a power law even at these finite temperature, because
we have prohibited spinons in our theory (equivalently there are no
defects in our height field $\chi$). This will be remedied below by
introducing gapped spinons. Second the form of the finite temperature
spatial correlations is roughly consistent with the naive expectation
that scaling should hold upto logarithmic corrections due to the
marginally irrelevant operator. What is interesting however is that
the important logarithmic correction occurs in the exponent $\sigma$
of the power-law (equal-time) spatial correlation - in its absence the
correlations are strictly local as shown explicitly in Section
\ref{spcorrsc}.  We also note that the correlator above does not
reproduce the zero temperature correlation function on simply
substituting $T=0$. This is only as expected as it arises entirely
from the irrelevant $u$ term. Note that while the scaling form for the
autocorrelation function was already obtained in the scaling limit,
one needs to include irrelevant operators to obtain finite spatial
correlations. Thus a marked asymmetry in the origin of spatial
and temporal correlations is evident.

It is useful to contrast the present model with other familiar models
right at their upper critical dimension - for instance the $O(N)$
quantum critical point in three spatial dimensions.  In all such cases
it is necessary to include irrelevant operators in order to obtain the
correct finite temperature correlations\cite{subir}. However in
contrast to other critical theories at their upper critical dimension
(eg. $\phi^4$ theory), the quantum Lifshitz transition fixed point has
operators with nontrivial scaling dimensions. The corresponding
correlation functions might have been expected to show scaling at
finite temperatures - but as we have seen the true behavior is more
intricate.

\subsubsection{Effect of Gapped Spinons}
\label{spnnsec}

Now consider introducing spinons - i.e. vortex defects in the height
field $\chi$ (these are absent in a pure quantum dimer model, but
are present in more physical representations of quantum magnets).
Assume that these spinons have an energy gap $E_c$ at the zero
temperature quantum critical point. Then, in the finite temperature
phase where an effective stiffness $\rho_{eff}$ is generated, in
addition to the core energy $E_c$, there is an additional
contribution to the energy that is logarithmically divergent with
the system size and takes the form: \be E_v=\frac{\rho_{eff}}{4\pi}
\log(L/a) \ee where $L$ is the system size, $a$ a microscopic
lengthscale at which we may ascribe to the system an effective
stiffness $\rho_{eff}$. Using the familiar Kosterlitz-Thouless
criterion, we conclude that the entropy of vortex production, which
is also logarithmically divergent with system size, wins over the
energy cost if $\frac{\rho_{eff}}{4\pi}<2T$. Since $\rho_{eff}/T$ goes to zero as $T \rightarrow 0$,
we conclude that the vortices
(spinons) will be in the plasma phase, and the correlator
(\ref{corr}) will eventually be an exponentially decaying function
with a decay length set by $\xi_{spinon}\propto
e^{-\frac{E_c}{2T}}$. Therefore, with a sufficiently large gap to
spinons we have three regimes. First, for $r \ll \xi_T$ we have the
quantum critical scaling regime with an effective system size cutoff
set by $\xi_T$. Next, for $\xi_T \ll r \ll \xi_{spinon}$ we have
power law correlators with a temperature dependent exponent $\sigma$.
Finally, for $r \gg \xi_{spinon}$ we have an exponentially decaying
function.

\section{Predictions for Numerical Experiments}
Numerical experiments on quantum dimer models could directly verify
the predictions of local criticality at these transitions. Perhaps
the most readily accessible case for numerical experiments is the
square lattice quantum dimer model with the RK hamiltonian. The zero
temperature phase transition between a zero tilt and the staggered
dimer phases that can be driven by varying parameters in the
Hamiltonian is known to be a highly fine tuned version of the
generic critical point discussed in \cite{huse,VBS}, and considered
in this paper. In particular it is known \cite{henley} that the
action (\ref{scalingaction}) describes the asymptotic properties of
the square lattice RK point with $K=\pi^2/4$ which may be compared
against exact results \cite{MEFisher}. Correlators of the monopole
insertion operator $V_r=\xi_r e^{i2\pi \chi_r}$  (where $\xi_r=\pm1,
\pm i$ is a Berry phase factor that oscillates on the four
sublattices) correspond to correlators of the dimer bond/plaquette
order. Since the bare RK point is a highly fine tuned critical
point, it lacks a bare quartic term ($u=0$), which can be seen from
the absence of logarithmic factors in the exact expressions for
correlation functions (which would otherwise arise if this
marginally irrelevant operator were present). Finite temperature
properties above the RK point are therefore expected to be as
follows. The  autocorrelation function of the monopole insertion
operators should obey $\omega/T$ scaling, with the (scaling function
part of the) spectral function given by equation (\ref{RKscalingfn})
and plotted in Fig. \ref{sfn}. The equal time correlators at
spatially separated points though should approach zero in the
scaling limit. Therefore the phenomenon of `local' criticality
should be visible in such numerical experiments. Non universal
corrections to scaling, as calculated above on inclusion of the
quartic term and gapped spinons, are not directly relevant to the
pure RK point, since, as we noted before, it is a fine tuned point
that lacks the quartic term, and the hard dimer constrain forbids
spinons. Hence, the corrections to scaling will arise from the least
irrelevant operator present, (e.g. the four monopole insertion
operator that is non-oscillating and hence appears in the coarse
grained action). A procedure similar to the one carried out here
with the quartic term for the generic case, needs to be repeated
with that operator to obtain the full asymptotics of the spatially
separated correlators.

\section{Conclusion}
In this paper we have studied certain aspects of the finite
temperature properties of the quantum Lifshitz transition discussed in
Ref. \onlinecite{huse,VBS,Ardonne}. We first obtained exact
information about the real time dynamical correlators at non-zero
temperatures of certain important physical operators in the scaling
limit. Such calculations are in general not possible for non-trivial
quantum critical points in dimensions bigger than one. The quantum
Lifshitz transition considered in this paper is a non-trivial quantum
phase transition that nevertheless admits a free Gaussian description
- this enables the calculation of the finite temperature dynamics.
One of the remarkable results of this calculation is that in the
scaling limit the correlators of the operators considered are strictly
local in space though they are non-trivial power laws in time. This
peculiar feature holds in the thermodynamic limit at non-zero
temperatures. On the other hand if the temperature is allowed to go to
zero first, and the thermodynamic limit taken later, spatial
dependences indeed arise even in the scaling limit.  Thus these
quantum transitions provide an explicit example of a certain kind of
`local' behavior at a quantum critical point. However we emphasize
that this is strictly a property of the model at non-zero
temperature. The zero temperature fixed point is described by a fairly
ordinary looking field theory.  Spatial dependence of the correlators
at non-zero temperature is restored once operators that are formally
irrelevant at the fixed point are included.  These corrections to
scaling were calculated for two different classes of irrelevant
perturbations, the quartic operator and gapped spinons, in Section \ref{beyondscaling}.

What lessons may we learn for other quantum critical points? First the
`local' structure of the finite temperature scaling found is
presumably special to this quantum Lifshitz transition - at least
within the class of bosonic quantum critical points that are
understood the best. It depends in part on realizing the special
circumstance of a theory at its upper critical dimension that
nevertheless has operators with nontrivial scaling
dimensions. Therefore, we expect that similar phenomena will {\em not}
arise even at the other non-trivial deconfined critical points studied
in Ref. \onlinecite{sc}. However in more complex situations with
fermionic degrees of freedom, quantum phase transition phenomena are
much less understood theoretically. The idea that some kind of spatial
locality may be associated with the quantum critical fluctuations has
been proposed at a phenomenological level to understand experiments in
a few materials (such as the cuprates or heavy fermions where gapless
fermionic excitations are undoubtedly present). Unfortunately it has
thus far not been possible to develop any serious theoretical
foundation for such ideas. Most of these proposals have assigned the
locality observed in the finite temperature quantum critical region,
to the local character of the zero temperature fixed point. In
contrast, the zero temperature fixed point studied in this paper is
{\em not} local in any sense. Rather, the local structure of
correlations only arises when we consider the thermodynamic limit of
the finite temperature system (and we ignore all irrelevant operators). Hence the
physics described in this paper provides a concrete example of a
possible alternate route to some kind of `local' criticality. It is hoped that the
mathematical structure of the model studied in this paper might help
in the search for similar phenomena in other models of
strongly correlated system.

\label{conc}
\section{Acknowledgements}
We would like to thank S. Sachdev for useful
discussions. A.V. would like to acknowledge support from the
Pappalardo Fellows Program at MIT and a Sloan Fellowship. TS is supported by the National Science Foundation
grant DMR-0308945,  the
NEC Corporation, the Alfred P. Sloan Foundation, and
The Research Corporation.


\begin{thebibliography}{x}
\bibitem{sc} T. Senthil, Ashvin Vishwanath, Leon Balents, Subir Sachdev and M. P.A. Fisher, Science {\bf 303}, 1490 (2004);
T. Senthil, L. Balents, S. Sachdev, A. Vishwanath, and M. P. A.
Fisher, Phys. Rev. B {\bf 70}, 144407 (2004).

\bibitem{huse} E. Fradkin, D.A. Huse, R. Moessner, V. Oganesyan, and
S. Sondhi, Phys. Rev. B {\bf 69}, 224415 (2004).

\bibitem{VBS} Ashvin Vishwanath, L. Balents, T. Senthil, Phys. Rev. B {\bf 69}, 224416 (2004).

\bibitem{henley} C. L. Henley J. Phys.: Condens. Matter {\bf 16} No.11 S891 (2004); C. L. Henley J. Stat. Phys. {\bf 89}, 483
(1997).

\bibitem{Ardonne} E. Ardonne, P. Fendley and E. Fradkin, Annals Phys. {\bf 310}, 493 (2004).





\bibitem{ReSaSuN} N.~Read and S.~Sachdev, Phys. Rev. Lett. {\bf 62}, 1694
  (1989);  N.~Read and S.~Sachdev, Phys. Rev. B {\bf 42}, 4568 (1990).

\bibitem{Frbook} E. Fradkin and S. A. Kivelson, Mod. Phys. Lett. B 4, 225 (1990);
E. Fradkin, {\em Field theories of Condensed Matter Systems}, Perseus Books (1991).

\bibitem{Grinstein} G.~Grinstein, Phys. Rev. B {\bf 23}, 4615 (1981).

\bibitem{varma} C.M. Varma, Phys Rev. B {\bf 55}, 14554 (1997).

\bibitem{si}Q. Si et al., Nature 413, 804 (2001) and Phys. Rev. B {\bf 68}, 115103 (2003).



\bibitem{sss} S. Sachdev, T. Senthil, R. Shankar
Phys. Rev. B {\bf 50}, 258 (1994).

\bibitem{MEFisher} M.~E.~Fisher and J.~Stephenson, Phys. Rev. 132, 1411
(1963).




\bibitem{subir} S. Sachdev, {\em Quantum Phase Transitions}, Cambridge
University Press (1999); S. Sachdev, Phys. Rev B {\bf 55}, 142
(1997).

\bibitem{footnotesum} We regulate the sum over $n$ with a factor
$e^{-\frac{n}{n_c}}$, to eliminate the logarithmic divergence, and
finally take $n_c\rightarrow \infty$.

\bibitem{footnote1} Note, that although the Fourier transform of
equation (\ref{auto-corr}) requires a short time cutoff to be defined for $\eta>1$, the expression for the
spectral function (\ref{A}) is valid for any $\eta>0$ and is
independent of this cutoff.


\end{thebibliography}
\end{document}